\documentclass[a4,10pt,twocolumn]{article}
\usepackage[left=2cm,right=2cm,top=2.1cm,bottom=2cm]{geometry}
\usepackage{graphicx} % standard LaTeX graphics tool
\usepackage{setspace}
\title{Einstein equations: exact solutions}
\date{}

\author{
Ji\v{r}\'{\i} Bi\v{c}\'{a}k\\
\small Institute of Theoretical Physics %\\
\small Charles University, Prague, Czech Republic \\
\small and \\
\small Albert Einstein Institute, Golm, Germany
}

\begin{document}

\maketitle

\section{Introduction}

Even in a linear theory like Maxwell's electrodynamics, in which
sufficiently general solutions of the field equations can be
obtained, one needs a good sample, a useful kit, of explicit exact
fields like the homogeneous field, the Coulomb monopole field, the
dipole, and other simple solutions, in order to gain a physical
intuition and understanding of the theory. In Einstein's general
relativity, with its nonlinear field equations, the discoveries
and analyses of various specific explicit solutions revealed most
of the unforeseen features of the theory. Studies of special
solutions stimulated questions relevant to more general
situations, and even after the formulation of a conjecture about a
general situation, newly discovered solutions can play a
significant role in verifying or modifying the conjecture. The
cosmic censorship conjecture assuming that ``singularities forming
in a realistic gravitational collapse are hidden inside horizons"
is a good illustration.

Albert Einstein presented the final version of his gravitational
field equations (EE) to the Prussian Academy in Berlin on November
18, 1915:
\begin{eqnarray}\label{I1}
R_{\mu\nu} - \frac{1}{2} g_{\mu\nu} R = \frac{8\pi G}{c^4} T_{\mu\nu}.
\end{eqnarray}
Here the spacetime metric tensor $g_{\mu\nu}(x^\rho)$, $\mu, \nu,
\rho, \ldots = 0, 1, 2, 3$, determines the invariant line element
$g = g_{\mu\nu} dx^\mu dx^\nu$, and acts also as a dynamical variable
describing the gravitational field; the Ricci tensor
$R_{\mu\nu}=g^{\rho\sigma} R_{\rho\mu\sigma\nu}$, where
$g^{\mu\rho}g_{\rho\nu} = \delta ^{\mu}_{\nu}$, is formed from the
Riemann curvature tensor $R_{\rho\mu\sigma\nu}$; both depend
nonlinearly on $g_{\alpha\beta}$ and $\partial _\mu
g_{\alpha\beta}$, and linearly on $\partial_\mu \partial_\nu
g_{\alpha\beta}$; the scalar curvature $R = g^{\mu\nu}
R_{\mu\nu}$. $T_{\mu\nu}(x^\rho)$ is the energy-momentum tensor of
matter (``sources''); and Newton's gravitational constant $G$ and
the velocity of light $c$ are fundamental constants. If not
stated otherwise, we use the geometrized units in which $G = c =
1$, and the same conventions as in \cite{MTW}, \cite{Wa}. For
example, in the case of perfect fluid with density $\rho$,
pressure $p$, and $4$-velocity $U^\mu$, $T_{\mu\nu}=(\rho+p) U_\mu
U_\nu + p g_{\mu\nu}$. To obtain a (local) solution of (\ref{I1})
in coordinate patch $\{x^\rho\}$ means to find ``physically
plausible'' (i.e., complying with one of the positive-energy conditions) functions $\rho(x^\rho)$, $p(x^\rho)$, $U_\mu(x^\rho)$,
and metric $g_{\mu\nu}(x^\rho)$ satisfying (\ref{I1}).
In vacuum $T_{\mu\nu}=0$ and (\ref{I1}) implies $R_{\mu\nu}=0$.

In 1917, Einstein generalized (\ref{I1}) by adding a cosmological
term $\Lambda g_{\mu\nu}$ ($\Lambda={\rm const}$):
\begin{eqnarray}\label{I2}
R_{\mu\nu} - \frac{1}{2} g_{\mu\nu} R + \Lambda g_{\mu\nu} = 8\pi T_{\mu\nu}~.
\end{eqnarray}
A homogeneous and isotropic static solution of (\ref{I2}) (with
metric (\ref{FLRW}), $k=+1$, $a={\rm const}$), in which the
``repulsive effect'' of $\Lambda > 0$ compensates the
gravitational attraction of incoherent dust (``uniformly
distributed galaxies'')---the Einstein static universe---marked
the birth of modern cosmology. Although it is unstable and lost
its observational relevance after the discovery of the expansion
of the Universe in the late 1920s, in 2004 a ``fine tuned''
cosmological scenario was suggested according to which our
universe starts asymptotically from an initial Einstein static
state and later enters an inflationary era, followed by a standard
expansion epoch [[see entry Cosmology: Mathematical Aspects]]. There are
many other examples of ``old'' solutions which turned out to act as
asymptotic states of more general classes of models.

\section{Invariant characterization and classification of the solutions}
\subsection{Algebraic classification}
The Riemann tensor can be decomposed as
\begin{equation}
R_{\alpha \beta \gamma \delta} = C_{\alpha \beta \gamma \delta} + E_{\alpha \beta \gamma \delta} + G_{\alpha \beta \gamma \delta},
\end{equation}
where $E$ and $G$ are constructed from $R_{\alpha \beta}, R$ and
$g_{\alpha \beta}$ (see, e.g., \cite{SK}); the \emph{Weyl
conformal tensor} $C_{\alpha \beta \gamma \delta}$ can be
considered as the ``characteristic of the pure gravitational
field" since, at a given point, it cannot be determined in terms
of the matter energy-momentum tensor $T_{\alpha \beta}$ (as $E$
and $G$ can using EE). Algebraic classification is based on
a classification of the Weyl tensor. This is best formulated
using 2-component spinors $\alpha_A$ ($A=1,2$), in terms of which
any Weyl spinor $\Psi_{ABCD}$ determining $C_{\alpha \beta \gamma
\delta}$ can be factorized:
\begin{equation}
\Psi_{ABCD} = \alpha_{(A} \beta_B \gamma_C \delta_{D)},
\end{equation}
brackets denote symmetrization; each of the spinors determines a
principal null direction, say, $k^\alpha = \alpha^A
\bar{\alpha}^{A'}$ [[see entry Spinors and Spin-coefficients]].
The Petrov-Penrose classification is based on coincidences among
these directions. A solution is of type \emph{I} (general case),
\emph{II, III}, and \emph{N} (``null") if all null directions are
different, or two, three, and all four coincide, respectively. It
is of type \emph{D} (``degenerate") if there are two double null
directions. The equivalent tensor equations are
simplest for type \emph{N}:
\begin{equation}\label{TypeN}
C_{\alpha \beta \gamma \delta} k^\delta = 0, \hspace{1mm} C_{\alpha \beta \gamma \delta}C^{\alpha \beta \gamma \delta} = 0, \hspace{1mm} C_{\alpha \beta \gamma \delta} C^{*\alpha \beta \gamma \delta} = 0,
\end{equation}
where $C^*_{\alpha \beta \gamma \delta} = \frac{1}{2}
\epsilon_{\alpha \beta \rho \sigma} {C^{\rho \sigma}}_{\gamma
\delta}$, $\epsilon$ is the Levi-Civita pseudotensor.
\subsection{Classification according to symmetries}
Most of the available solutions have some exact continuous symmetries
which preserve the metric. The corresponding group of motions is
characterized by the number and properties of its Killing vectors
$\xi^\alpha$ satisfying the Killing equation $(\pounds_\xi
g)_{\alpha \beta} = \xi_{\alpha ; \beta} + \xi_{\beta ; \alpha} =
0$ ($\pounds$ is the Lie derivative) and by the nature (spacelike,
timelike or null) of the group orbits. For example, axisymmetric,
stationary fields possess two commuting Killing vectors, of
which one is timelike. Orbits of the axial Killing vector are
closed spacelike curves of finite length, which vanishes at the
axis of symmetry. In cylindrical symmetry, there exist two
spacelike commuting Killing vectors. In both cases, the vectors
generate a 2-dimensional Abelian group. The 2-dimensional group
orbits are timelike in the stationary case and spacelike in the
cylindrical symmetry.

If a timelike $\xi^\alpha$ is hypersurface-orthogonal, $\xi_\alpha
= \lambda \Phi_{, \alpha}$ for some scalar functions $\lambda,
\Phi$, the spacetime is \emph{static}. In coordinates with $\xi =
\partial_t$, the metric is
\begin{equation}
g = - e^{2U} dt^2 + e^{-2U}\gamma_{ik} dx^i dx^k,
\end{equation}
where $U, \gamma_{ik}$ do not depend on $t$. In vacuum, $U$
satisfies the potential equation $U_{:a}^{:a} = 0$, the covariant
derivatives (denoted by :) are w.r.t. the 3-dimensional metric $\gamma_{ik}$. A
classical result of Lichnerowicz states that if the vacuum metric
is smooth everywhere and $U \rightarrow 0$ at infinity, the
spacetime is flat (for refinements, see M.T.
Anderson, 2000).

In cosmology, we are interested in groups whose regions of
transitivity (points can be carried into one another by symmetry
operations) are 3-dimensional spacelike hypersurfaces (homogeneous but anisotropic models of the Universe). The
3-dimensional \emph{simply} transitive groups $G_3$ were
classified by Bianchi in 1897 according to the possible distinct
sets of structure constants but their importance in cosmology
was discovered only in the 1950's. There are nine types: Bianchi I
to Bianchi IX models. The line element of the Bianchi universes
can be expressed in the form
\begin{equation}
g = - dt^2 + g_{ab}(t) \omega^a \omega^b,
\end{equation}
where the time-independent 1-forms $\omega^a = E_\alpha^a
dx^\alpha$ satisfy the relations $d\omega^\alpha = - \frac{1}{2}
C^a_{bc} \omega^b \wedge \omega^c$, \emph{d} is the exterior
derivative and $C^a_{bc}$ are the structure constants [[see entry
Cosmology: Mathematical Aspects for more details]].

The standard Friedmann-Lemaitre-Robertson-Walker (FLRW) models
admit in addition an isotropy group SO(3) at each point. They can
be represented by the metric
\begin{equation}\label{FLRW}
\!g\! =\!- dt^2 + [a(t)]^2\left( \frac{dr^2}{1-kr^2} + r^2 (d\theta^2 + \sin^2\theta d\varphi^2)\right),
\end{equation}
in which $a(t)$, the ``expansion factor", is determined by matter
via EE, the curvature index $k = -1, 0, +1$, the 3-dimensional
spaces $t=$ const have a constant curvature $K= k/a^2; r \in [0,1]$
for closed ($k=+1$) universe, $r \in [0,\infty)$ in open
($k=0,-1$) universes (for another description, [[see entry
Cosmology: Mathematical Aspects]]).

There are 4-dimensional spacetimes of constant curvature solving
EE (\ref{I2}) with $T_{\mu \nu} = 0$: the Minkowski, de Sitter,
and anti de Sitter spacetimes. They admit the same number (10) of
independent Killing vectors, but interpretations of the
corresponding symmetries differ for each spacetime.

If $\xi^\alpha$ satisfies $\pounds_\xi g_{\alpha \beta} = 2 \Phi
g_{\alpha \beta}, \Phi=$const, it is called a homothetic (Killing)
vector. Solutions with proper homothetic motions $\Phi \not = 0$
are ``self-similar". They cannot in general be asymptotically flat
or spatially compact but can represent asymptotic states of more
general solutions. In \cite{SK}, a summary of solutions with proper
homotheties is given; their role in cosmology is analyzed by J.
Wainwright and G.F.R. Ellis (eds) 1996, and by A.A. Coley, 2003;
for mathematical aspects of symmetries in general relativity, see
G.S. Hall, 2004.

There are other schemes for invariant classification of exact
solutions (reviewed in \cite{SK}): the algebraic
classification of the Ricci tensor and energy-momentum tensor of
matter; the existence and properties of preferred vector
fields and corresponding congruences; local isometric embeddings
into flat pseudo-Euclidian spaces, etc.
\section{Minkowski (M), de Sitter (dS), and anti de Sitter (AdS) spacetimes}
These metrics of constant (zero, positive, negative) curvature are
the simplest solutions of (\ref{I2}) with $T_{\mu \nu}=0$ and
$\Lambda=0, \Lambda>0, \Lambda<0$, respectively. The standard
topology of M is $R^4$. The dS has the topology $R^1 \times S^3$
and is best represented as a 4-dimensional hyperboloid $-v^2 + w^2
+x^2 +y^2 +z^2 = (3/\Lambda)$ in a 5-dimensional flat space with
metric $g= -dv^2 +dw^2 +dx^2 + dy^2 +dz^2$. The AdS has the
topology $S^1 \times R^3$; it is a 4-dimensional hyperboloid
$-v^2 - w^2 +x^2 +y^2 +z^2 = -(3/\Lambda), \Lambda<0$, in flat 5-d
space with signature $(-,-,+,+,+)$. By unwrapping the circle $S^1$
and considering the universal covering space, one gets rid of
closed timelike lines.

These spacetimes are all conformally flat and can be conformally mapped into
portions of the Einstein universe [[see entry Asymptotic Structure and
Conformal Infinity]]. However, their conformal structure is
globally different. In M one can go to infinity along
timelike/null/spacelike geodesics and reach five qualitatively
different sets of points: future/past timelike infinity
\textit{i}$^\pm$, future/past null infinity $\mathcal{I}^\pm$, and
spacelike infinity \textit{i}$^0$. In dS, there are only past and
future conformal infinities $\mathcal{I}^-, \mathcal{I}^+$, both
being \emph{spacelike} (on the Einstein cylinder, the de Sitter
spacetime is a ``horizontal strip" with
$\mathcal{I}^+/\mathcal{I}^-$ as the ``upper/lower circle"). The
conformal infinity in AdS is timelike.

As a consequence of spacelike $\mathcal{I}^\pm$ in dS, there exist
both particle (cosmological) and event horizons for geodesic
observers \cite{HE}. dS plays a (doubly) fundamental role in the
present-day cosmology: it is an approximate model for inflationary
paradigm near the Big Bang and it is also the asymptotic state (at
$t \rightarrow \infty$) of cosmological models with a positive
cosmological constant. Since recent observations indicate that
$\Lambda >0$, it appears to describe the future state of our
Universe. AdS has come recently to the fore due to the ``holographic"
conjecture [[see entry AdS/CFT Correspondence]].

Christodoulou and Klainermann, and Friedrich proved that M, dS,
and AdS are \emph{stable} with respect to general,
\emph{nonlinear} (though ``weak") \emph{vacuum}
perturbations---result not known for any other solution of
EE [[see entry Stability of Minkowski Space]].
\section{Schwarzschild and Reissner-Nordstr\"{o}m metrics}
These are spherically symmetric spacetimes---the SO$_3$ rotation group acts on them as an isometry group with spacelike, 2-dimensional orbits. The metric can be brought into the form
\begin{equation}
g = - e^{2\nu} dt^2 + e^{2\lambda} dr^2 + r^2 (d\theta^2 + \sin^2\theta d\varphi^2),
\end{equation}
$\nu(t,r), \lambda(t,r)$ must be determined from EE. In vacuum, we are led uniquely to the Schwarzschild metric
\begin{equation}\label{Schwarzschild}
g \!=\! -\!\left(\! 1\!-\! \frac{2M} {r} \!\right) dt^2 
\!+\! \left(\! 1\!-\! \frac{2M} {r} \!\right)^{-1} \!\!\!\!dr^2 \!+\! r^2 (d\theta^2 + \sin^2\theta d\varphi^2),
\end{equation}
where $M=$ const has to be interpreted as mass, as test particle orbits show. The spacetime is static at $r>2M$, i.e., outside the Schwarzschild radius at $r=2M$, and asymptotically ($r \rightarrow \infty$) flat.

Metric (\ref{Schwarzschild}) describes the exterior gravitational
field of an arbitrary (static, oscillating, collapsing or
expanding) spherically symmetric body (spherically symmetric
gravitational waves do not exist). It is the most influential
solution of EE. The essential tests of general
relativity---perihelion advance of Mercury, deflection of both
optical and radio waves by the Sun, and signal retardation---are
based on (\ref{Schwarzschild}) or rather on its expansion in
$M/r$. Space missions have been proposed that could lead to
measurements of ``post-post-Newtonian" effects [[see entry
General Relativity: Experimental Tests  and \cite{MTW}]]. The full
Schwarzschild metric is of importance in astrophysical processes
involving compact stars and black holes.

\begin{figure}[b]
\begin{center}
\includegraphics[width=.44\textwidth]{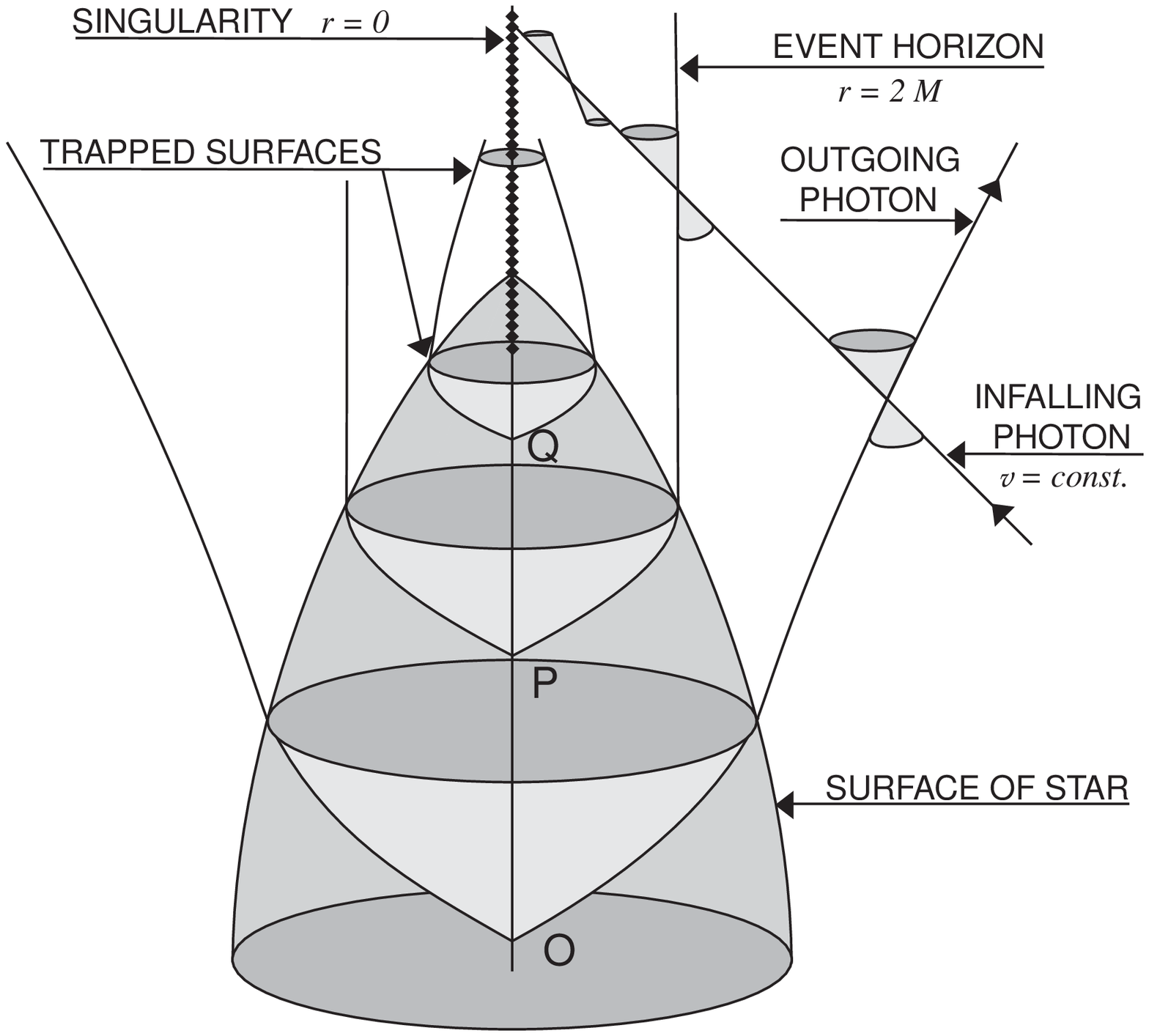}
\end{center} \small 
{\bf Figure 1} Gravitational collapse of a spherical star (the interior
of the star is shaded). The light cones of three events,
$O$, $P$, $Q$, at the center of the star, and of three events
outside the star are illustrated. The event horizon,
the trapped surfaces, and the singularity formed
during the collapse are also shown. Although the singularity appears
to lie along the direction of time, from the character of the
light cone outside the star but inside
the event horizon we can see that it has a spacelike character.
\end{figure}
%\end{center}

Metric (\ref{Schwarzschild}) describes the spacetime outside a
spherical body collapsing through $r=2M$ into a spherical black
hole. In Fig. 1, the formation of an event horizon and trapped
surfaces is indicated in ingoing Eddington-Finkelstein coordinates
$(v,r, \theta, \varphi)$ where $v=t+r+2M \log (r/2M -1)$ so that
$(v, \theta, \varphi)=$ const are ingoing radial null geodesics.
The interior of the star is described by another metric (e.g., the
Oppenheimer-Snyder collapsing dust solution---see below). The
Kruskal extension of the Schwarzschild solution, its
compactification, the concept of the bifurcate Killing horizon,
etc., are analyzed in [[entry Stationary Black Holes]] and in
\cite{MTW}, \cite{HE}, {\cite{JiBi}.

%\begin{center} 

The \emph{Reissner-Nordstr\"{o}m solution} describes the exterior
gravitational and electromagnetic fields of a spherical body with
mass $M$ and charge $Q$. The energy-momentum tensor on the right-hand side of EE is that 
of the electromagnetic field produced by the charge; the field satisfies the curved-space
Maxwell equations. The metric reads
\begin{eqnarray}\label{Reissner-Nordstrom}
g = &-& \left( 1- \frac{2M} {r} + \frac{Q^2}{r^2} \right) dt^2 +\\\nonumber
&+& \left( 1- \frac{2M} {r} + \frac{Q^2}{r^2} \right)^{-1}\!\!\! dr^2 
+ r^2 (d\theta^2 + \sin^2\theta d\varphi^2).
\end{eqnarray}

\begin{figure*}[ht]
\begin{center}
\includegraphics[width=.75\textwidth]{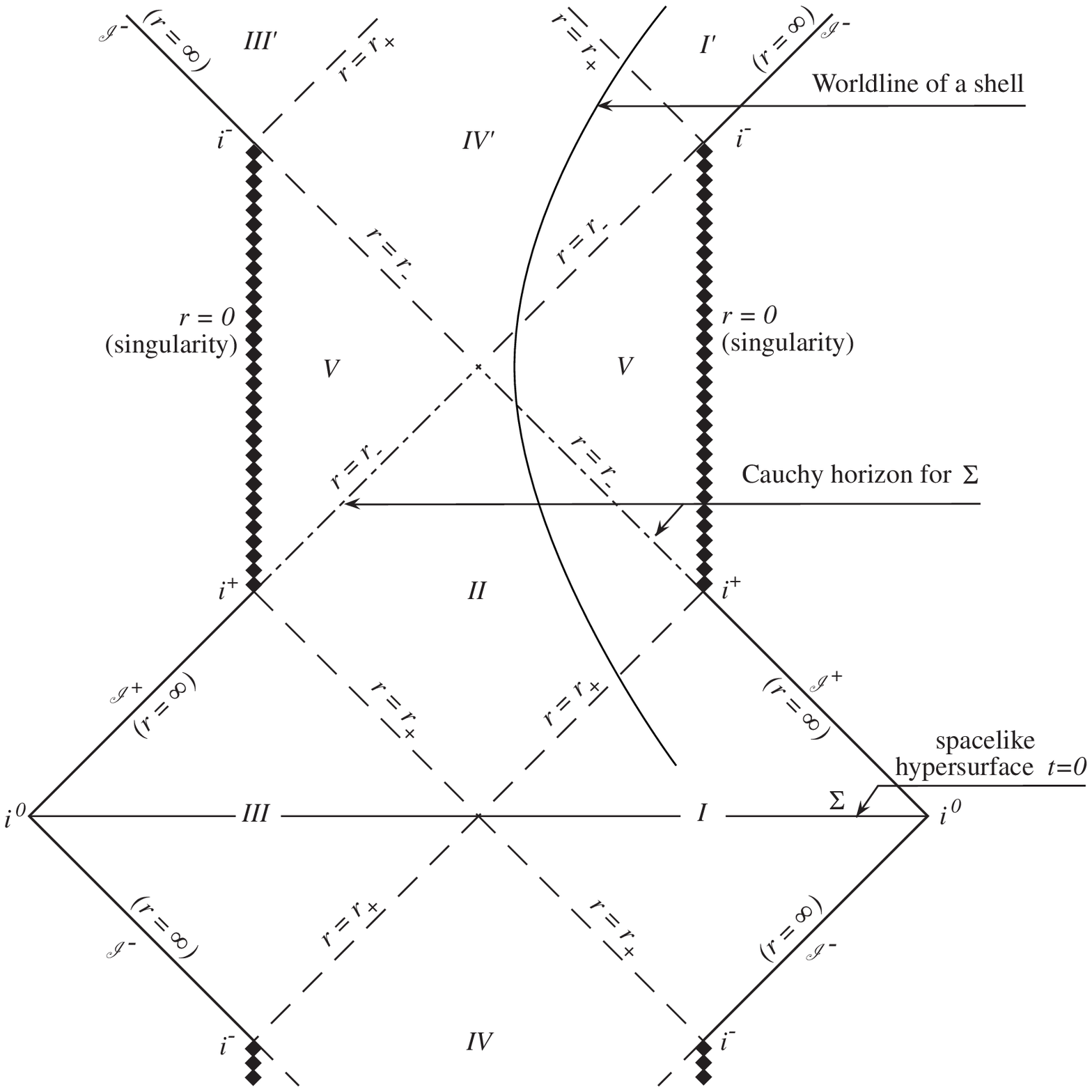}
\end{center} \small 
{\bf Figure 2} ~~The compactified Reissner-Nordstr\"{o}m spacetime
representing a non-extreme black hole consists of
an indefinite chain of asymptotic regions (``universes'')
connected by ``wormholes'' between timelike singularities.
The worldline of a shell collapsing from ``universe'' $I$ and
re-emerging in ``universe'' $I'$ is indicated. The inner horizon
at $r=r_-$ is the Cauchy horizon for a spacelike hypersurface $\Sigma$.
It is unstable and thus it will very likely prevent such a process.
\end{figure*}

The analytic extension of the electrovacuum metric
(\ref{Reissner-Nordstrom}) is qualitatively different from the
Kruskal extension of the Schwarzschild metric. In the case
$Q^2>M^2$ there is a \emph{naked singularity} (visible from $r
\rightarrow \infty$) at $r=0$ where curvature invariants diverge.
If $Q^2<M^2$, the metric describes a (generic) static charged
black hole with two event horizons at $r=r_\pm = M \pm (M^2 -
Q^2)^{1/2}$. The Killing vector $\partial / \partial t$ is null at
the horizons, timelike at $r>r_+$ and $r<r_-$, but spacelike
between the horizons. The character of the extended spacetime is
best seen in the compactified form, Fig. 2, in which worldlines
of radial light rays are 45-degree lines. Again, two infinities
(right and left, in regions I and III) arise (as in the
Kruskal-Schwarzschild diagram---see [[entry Stationary Black
Holes]]), however, the maximally extended geometry consists of an
infinite chain of asymptotically flat regions connected by
``wormholes" between the singularities at $r=0$. In contrast to
the Schwarzschild singularity, the singularities are
timelike---they do not block the way to the future. The inner
horizon $r=r_-$ represents a \emph{Cauchy horizon} for a typical
initial hypersurface like $\Sigma$ (Fig. 2): What is happening in
regions V is in general influenced not only by data on $\Sigma$
but also at the singularities. The Cauchy horizon is unstable (for
Refs., see \cite{JiBi} and recent work by M. Dafermos (2005)).

For $M^2=Q^2$ the two horizons coincide at $r_+ = r_- = M$. Metric
(\ref{Reissner-Nordstrom}) describes \emph{extreme}
Reissner-Nordstr\"{o}m \emph{black holes}. The horizon becomes
degenerate and its surface gravity vanishes [[see entry Stationary
Black Holes]]. Extreme black holes play a significant role in
string theory \cite{Ort}.

\section{Stationary axisymmetric solutions}
Assume the existence of two commuting Killing vectors---timelike
$\xi^\alpha$ and axial $\eta^\alpha$ ($\xi^\alpha\xi_\alpha<0,
\eta^\alpha\eta_\alpha>0$), $\xi^\alpha$ normalized at
(asymptotically flat) infinity, $\eta^\alpha$ at the rotation
axis. They generate 2-dimensional orbits of the group $G_2$.
Assume there exist 2-spaces orthogonal to these orbits. This is
true in vacuum and also in case of electromagnetic fields or perfect
fluids whose 4-current or 4-velocity lies in the surfaces of
transitivity of $G_2$ (e.g., toroidal magnetic fields are
excluded). The metric can then be written in Weyl's coordinates
$(t, \rho, \varphi, z)$
\begin{equation}\label{Weyl}
g = - e^{2U}(dt + Ad\varphi)^2 + e^{-2U}[e^{2k}(d\rho^2+dz^2)+\rho^2d\varphi^2],
\end{equation}
$U$, $k$, and $A$ are functions of $\rho, z$.

The most celebrated vacuum solution of the form (\ref{Weyl}) is
the \emph{Kerr metric} for which $U, k, A$ are ratios of simple
polynomials in spheroidal coordinates (simply related to $(\rho,
z)$). The Kerr solution is characterized by mass $M$ and specific
angular momentum $a$. For $a^2>M^2$, it describes an asymptotically
flat spacetime with a naked singularity. For $a^2 \leq M^2$, it
represents a \emph{rotating black hole} that has two horizons
which coalesce into a degenerate horizon for $a^2=M^2$---an
\emph{extreme Kerr black hole}. The two horizons are located at
$r_\pm = M \pm (M^2-a^2)^{\frac{1}{2}}$ ($r$ being the Boyer-Lindquist
coordinate---[[see entry Stationary Black Holes]]). As with the
Reissner-Nordstr\"{o}m black hole, the singularity inside is
timelike and the inner horizon is an (unstable) Cauchy horizon.
The analytic extension of the Kerr metric resembles Fig. 2 (see
\cite{FN}, \cite{HE}, \cite{MTW}-\cite{Wa} for details).

Thanks to the black-hole uniqueness theorems [[see entry
Stationary Black Holes]], the Kerr metric is the unique solution
describing all rotating black holes in vacuum. If the cosmic
censorship conjecture holds, Kerr black holes represent the end
states of gravitational collapse of astronomical objects with
supercritical masses. According to prevalent views, they reside in
the nuclei of most galaxies. Unlike with a spherical collapse,
there are no exact solutions available which would represent the
formation of a Kerr black hole. However, starting from metric
(\ref{Weyl}) and identifying, e.g., $z=b=$ const and $z=-b$ (with the region $-b<z<b$ being cut off), one can
construct thin material disks which are physically plausible and
can be the sources of the Kerr metric even for $a^2>M^2$ (see
\cite{JiBi} for details).

In a general case of metric (\ref{Weyl}), EE in vacuum imply the
\emph{Ernst equation} for a complex function $f$ of $\rho$ and $z$
:
\begin{equation}\label{Ernst}
(\Re f)[f_{,\rho\rho} + f_{,zz} + \frac{1}{\rho} f_{,\rho}] = f_{,\rho}^2 + f_{,z}^2,
\end{equation}
or, equivalently, $(\Re f) \triangle f = (\nabla f)^2$, where
$f=e^{2U}+ib$, $U$ enters (\ref{Weyl}), and $b(\rho,z)$ is a
``potential" for $A(\rho,z)$: $A_{,\rho} = \rho e^{-4U}b_{,z}$,
$A_{,z} = -\rho e^{-4U}b_{,\rho}$; $k(\rho,z)$ in (\ref{Weyl}) can
be determined from $U$ and $b$ by quadratures. Tomimatsu and Sato
(TS) exploited symmetries of (\ref{Ernst}) to construct metrics
generalizing the Kerr metric. Replacing $f$ by $\xi =
(1-f)/(1+f)$, one finds that in case of the Kerr metric $\xi^{-1}$ is a linear function in
the prolate spheroidal coordinates,
whereas for TS solutions $\xi$ is a quotient of higher-order
polynomials. A number of other solutions of Eq.
(\ref{Ernst}) were found but they are of lower significance than
the Kerr solution (cf. Ch. 20 in \cite{SK}).

These solutions inspired ``\emph{solution generating methods}" in
general relativity. The Ernst equation can be regarded as the
integrability condition of a system of linear differential
equations. The problem of solving such a system can be
reformulated as the Riemann-Hilbert problem in complex function
theory [[see entries Riemann-Hilbert problem, Integrable
Systems]]. We refer to \cite{SK} and \cite{BV} where these
techniques using B\"{a}cklund transformations, inverse scattering
method, etc. are also applied in the nonstationary context of two
spacelike Killing vectors (waves, cosmology). In the stationary case,
all asymptotically flat, stationary, axisymmetric vacuum solutions
can, in principle, be generated. It is known how to generate
fields with given values of multipole moments, though the required
calculations are staggering. By solving the Riemann-Hilbert
problem with appropriate boundary data, Neugebauer and Meinel
constructed the exact solution representing a rigidly rotating
thin disk of dust (cf. \cite{SK}, \cite{JiBi}).

A subclass of metrics (\ref{Weyl}) is formed by \emph{static Weyl
solutions} with $A=b=0$. Eq. (\ref{Ernst}) then becomes the Laplace
equation $\Delta U=0$. The non-linearity of EE enters only the
equations for $k$: $k_{,\rho} = \rho (U_{,\rho}^2 - U_{,z}^2)$,
$k_{,z} = 2 \rho U_{,\rho} U_{,z}$. The class contains some
explicit solutions of interest: the ``linear superposition" of
collinear particles with string-like singularities between them
which keep the system in static equilibrium; solutions
representing external fields of counter-rotating disks, e.g.,
those which are ``inspired" by galactic Newtonian potentials;
disks around black holes and some other special solutions
\cite{SK}, \cite{Bo1}, \cite{JiBi}, \cite{OS}.

There are solutions of the Einstein-Maxwell equations representing
external fields of masses endowed with electric charges, magnetic
dipole moments, etc. \cite{SK}. Best known is the
\emph{Kerr-Newman metric} characterized by parameters $M, a$, and
charge $Q$. For $M^2 \geq a^2 + Q^2$ it describes a \emph{charged,
rotating black hole}. Owing to the rotation, the charged black
hole produces also a magnetic field of a dipole type. All the
\emph{black hole} solutions can be generalized to \emph{include a
non-vanishing} $\Lambda$ (for various applications, see
\cite{OS}). Other generalizations incorporate the so-called NUT
(Newman-Unti-Tamburino) parameter (corresponding to a ``gravomagnetic monopole") or an
``external" magnetic/electric field or a parameter leading to
``uniform" acceleration (see \cite{SK}, \cite{JiBi}). Much
interest has recently been paid to black-hole (and other)
solutions with various types of gauge fields and to
multi-dimensional solutions. Refs. \cite{FN} and \cite{Ort} are
two examples of good reviews.

\begin{figure}[b]
\begin{center}
\includegraphics[width=.5\textwidth]{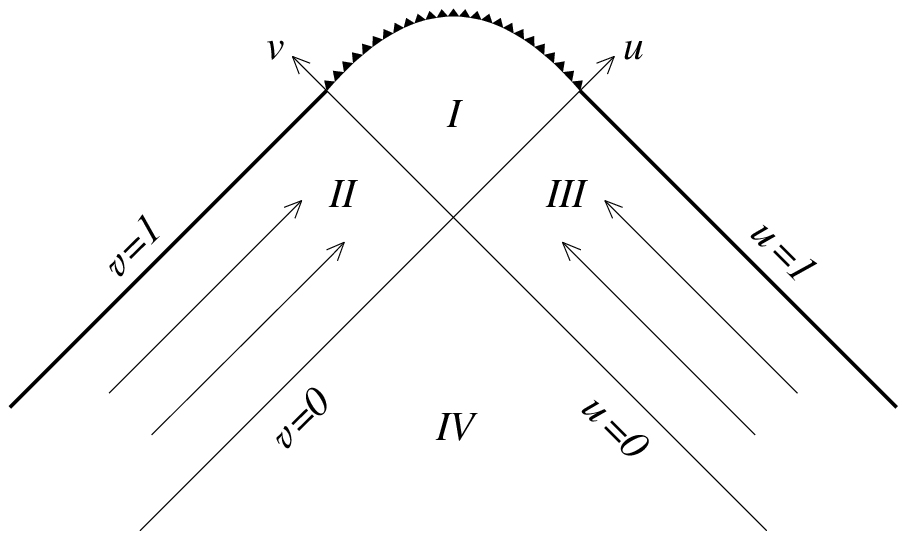}
\end{center} \small 
{\bf Figure 3}~~ A spacetime diagram indicating a collision of two plane-fronted
gravitational waves which come from regions {\it II} and {\it III},
collide in region $I$, and produce a spacelike singularity.
Region {\it IV} is flat.
\end{figure}

\section{Radiative solutions}
\subsection{Plane waves and their collisions}

The best known class are ``plane-fronted gravitational waves with
parallel rays'' (pp-waves) which are defined by the condition that
the spacetime admits a covariantly constant \emph{null} vector
field $k^\alpha$: $k_{\alpha;\beta} = 0$. In suitable null
coordinates $u,v$ such that $k_\alpha=u_{,\alpha}$,
$k^\alpha=\left(\partial/\partial v\right)^\alpha$, and complex
$\zeta$ which spans the wave $2$-surfaces $u={\rm const}$, $v={\rm
const}$ with Euclidean geometry, the metric reads
\begin{equation}
\label{RS1}
g = 2d \zeta d \bar \zeta - 2 du dv - 2H(u, \zeta, \bar
\zeta) du^2,
\end{equation}
$H(u, \zeta, \bar\zeta)$ is a real function. The vacuum EE imply
$H_{,\zeta \bar\zeta}=0$ so that $2H = f(u,\zeta)+\bar
f(u,\bar\zeta)$, $f$ is an arbitrary function of $u$, analytic in
$\zeta$. The Weyl tensor satisfies equations (\ref{TypeN})---the field is of type \emph{N} as is the field of plane
electromagnetic waves. In the null tetrad $\{ k^\alpha,l^\alpha,
m^\alpha {\rm (complex)} \}$ with $l^\alpha k_\alpha = -1,
m^\alpha \bar m_\alpha = 1 $, all other products vanishing, the
only nonzero projection of the Weyl tensor, $\Psi =
C_{\alpha\beta\gamma\delta} l^\alpha \bar m^\beta l^\gamma
m^\delta = H_{,\bar\zeta \bar\zeta}$, describes the
\emph{transverse} component of a wave propagating in the
$k^\alpha$ direction. Writing $\Psi = {\cal A} e^{i\Theta}$, the
real ${\cal A}>0$ is the \emph{amplitude} of the wave, $\Theta$
describes \emph{polarization}. Waves with $\Theta={\rm const}$
are called linearly polarized. Considering their effect on test
particles one finds plane waves are transverse.

The simplest waves are homogeneous in the sense that
$\Psi$ is constant along the wave surfaces. One gets $f(u,\zeta) =
{1\over 2}{\cal A}(u) \; e^{i\Theta(u)} \zeta^2$. Instructive are
\emph{sandwich waves}, e.g., waves with a ``square profile'':
${\cal A} =0$ for $u<0$ and $u>a^2$, ${\cal A} =a^{-2}={\rm
const}$ for $0\le u\le a^2$. This example demonstrates, within
\emph{exact} theory, that the waves travel with the speed of
light, produce relative accelerations of test particles, focus
astigmatically generally propagating parallel rays, etc. The
focusing effects have a remarkable consequence: there exists no 
global spacelike hypersurface on which initial data could be specified---plane wave spacetimes contain no global Cauchy hypersurface.

\emph{Impulsive} plane waves can be generated by boosting a
``particle'' at rest to the velocity of light by an appropriate
limiting procedure. The ultrarelativistic limit of, e.g., the
Schwarzschild metric (the so-called Aichelburg-Sexl solution) can be employed
as a ``limiting incoming state'' in black hole encounters (cf.
monograph by P.D. d'Eath, 1996). Plane-fronted waves have been
used in quantum field theory. For a review of exact impulsive
waves, see \cite{OS}.

A \emph{collision of plane waves} represents an exceptional
situation of nonlinear wave interactions which can be analyzed
exactly. Fig. 3 illustrates a typical case in which the collision
produces a spacelike singularity. The initial value problem with
data given at $v=0$ and $u=0$ can be formulated in terms of the
equivalent matrix Riemann-Hilbert problem [[see entry
Riemann-Hilbert problem]]; it is related to the hyperbolic
counterpart of the Ernst equation (\ref{Ernst}). For reviews, see
\cite{JG}, \cite{SK}, and \cite{JiBi}.

\subsection{Cylindrical waves}

Discovered by G. Beck in 1925 and known today as the Einstein-Rosen waves
(1937), these vacuum solutions helped to clarify a number
of issues, such as energy loss due to the waves, asymptotic structure of radiative spacetimes, dispersion of waves, quasi-local mass-energy, cosmic censorship conjecture, or quantum gravity
in the context of midisuperspaces (see \cite{JiBi}, \cite{BV}).

In the metric
\begin{equation}
\label{ds4cyl}
g = e^{2(\gamma - \psi)}(-dt^2+d\rho ^2)+e^{2\psi} dz^2+\rho ^2
e^{-2\psi} d\varphi^2,
\end{equation}
$\psi(t,\rho)$ satisfies the flat-space wave equation and
$\gamma(\rho,t)$ is given in terms of $\psi$ by quadratures. Admitting
a ``cross term'' $\sim \omega(t,\rho)\; dz \;d\phi$, one acquires a second degree of freedom
(a second polarization) which makes all field equations nonlinear.

\subsection{Boost-rotation symmetric spacetimes}

These are the only explicit solutions available which are
radiative and represent the field of finite sources. Fig. 4 shows
two particles uniformly accelerated in opposite directions. In the
space diagram (left), the ``string'' connecting the particles is
the ``cause'' of the acceleration. In ``Cartesian-type''
coordinates and the $z$-axis chosen as the symmetry axis, the boost
Killing vector has a flat-space form, $\zeta = z (\partial/\partial
t) + t  (\partial/\partial z)$, the same is true for the axial Killing
vector. The metric contains two functions of variables
$\rho^2\equiv x^2+y^2$ and $\beta^2\equiv z^2-t^2$. One satisfies the
flat-space wave equation, the other is determined by quadratures.

\begin{figure}[t]
\begin{center}
\includegraphics[width=.48\textwidth]{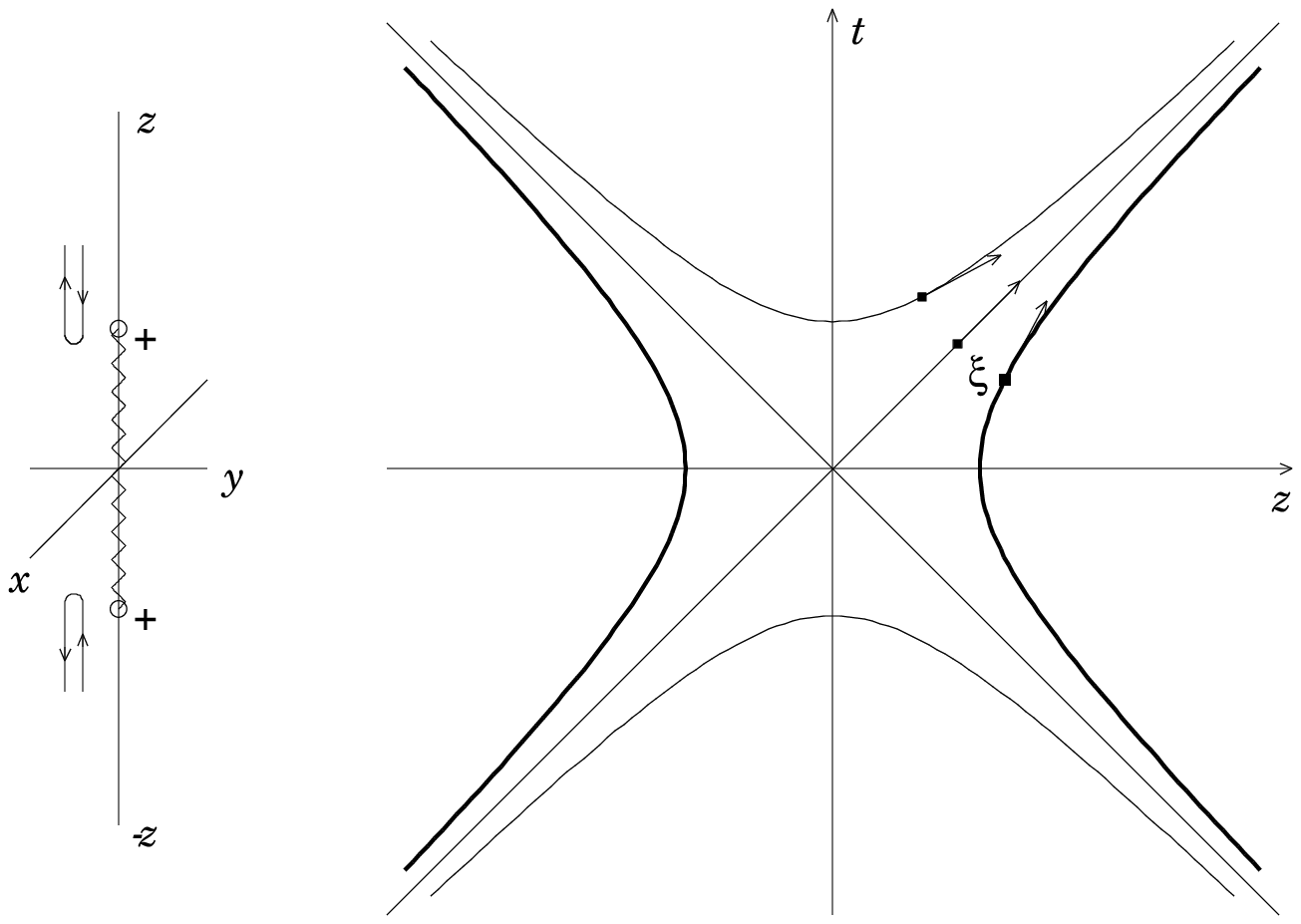}
\end{center} \small 
{\bf Figure 4}~~ Two particles uniformly accelerated in opposite directions.
Orbits of the boost Killing vector (thinner hyperbolas) are spacelike
in the region $t^2>z^2$.
\end{figure}

The unique role of these solutions is exhibited by the theorem
which states that in axially symmetric, locally asymptotically
flat spacetimes in the sense that a null infinity [[see entry
Asymptotic structure and conformal infinity]] exists but not
necessarily globally, the only \emph{additional} symmetry that
does not exclude gravitational radiation is the \emph{boost}
symmetry. Various radiation characteristics can be expressed
explicitly in these spacetimes. They have been used as tests in
numerical relativity and approximation methods. The best known
example is the C-metric (representing accelerating black holes, in
general charged and rotating, and admitting $\Lambda$)---see
\cite{Bo2}, \cite{JiBi}, \cite{SK}, \cite{OS}.

\subsection{Robinson-Trautman solutions}

These solutions are algebraically special but in general they do
\emph{not} possess any symmetry. They are governed by a function
$P(u,\zeta,\bar\zeta)$ ($u$ is the retarded time, $\zeta$ a complex
spatial coordinate) which satisfies a 4th-order nonlinear
parabolic differential equation.  Studies by Chru\'{s}ciel and
others have shown that RT solutions of Petrov type II exist
globally for all positive ``times'' $u$ and converge
asymptotically to a Schwarzschild metric, though the extension
across the ``Schwarzschild-like'' horizon can only be made with a
finite degree of smoothness. Generalization to the cases with
$\Lambda>0$ gives explicit models supporting the cosmic
no-hair conjecture (an exponentially fast approach to the de
Sitter spacetime) under the presence of gravitational waves. See
\cite{Bo2}, \cite{JiBi}, \cite{SK}.

\section{Material sources}

Finding physically sound material sources in an analytic form even for some simple vacuum metrics remains an open problem. Nevertheless, there are solutions representing regions of spacetimes filled with matter which are of considerable interest.

One of the simplest, the spherically symmetric Schwarzschild interior solution with incompressible fluid as its source, represents ``a star'' of uniform density, $\rho = \rho_0 =$ const:
\begin{eqnarray}\label{M1}\nonumber
g = &-& \left[ \frac{3}{2} \sqrt{1-AR^2} - \frac{1}{2} \sqrt{1 - Ar^2} \right]^2 dt^2 +\\
&+& \frac{dr^2}{1-Ar^2} + r^2 (d\theta^2 + \sin^2 \theta d \varphi^2),
\end{eqnarray}
$A = 8\pi \rho_0 /3 =$ const, $R$ is the radius of the star.

The equation of hydrostatic equilibrium yields pressure inside the star:
\begin{eqnarray}\label{M2}
8\pi p = 2A \frac{\sqrt{1-Ar^2} - \sqrt{1-AR^2}}{3\sqrt{1-AR^2} - \sqrt{1-Ar^2}}.
\end{eqnarray}
Solution (\ref{M1}) can be matched at $r=R$, where $p = 0$, to the
exterior vacuum Schwarzschild solution (\ref{Schwarzschild}) if
the Schwarzschild mass $M = \frac{1}{2}AR^3$. Although
``incompressible fluid'' implies an infinite speed of sound, the
above solution provides an instructive model of relativistic
hydrostatics. A Newtonian star of uniform density can have an
arbitrarily large radius $R = \sqrt{3p_c/2\pi\rho_0^2}$ and mass $M
= (p_c/\rho^2_0)\sqrt{6p_c/\pi}$, $p_c$ is the central pressure.
However, (\ref{M2}) implies that (i) $M$ and $R$ satisfy the
inequality $2M/R \leq 8/9$, (ii) equality is reached as $p_c$
becomes infinite and $R$ and $M$ attain their limiting values
$R_{\mbox{\tiny{lim}}} = (3\pi \rho_0)^{-\frac{1}{2}} =
(9/4)M_{\mbox{\tiny{lim}}}$. For a density typical in neutron
stars, $\rho _0 = 10^{15}$ g cm$^{-3}$, we get
$M_{\mbox{\tiny{lim}}} \doteq 3.96 M_{\odot}$ ($M _{\odot}$ solar
mass)---even this simple model shows that in Einstein's theory
neutron stars can only be a few solar masses. In addition, one can
prove that the ``Buchdahl's inequality" $2M/R \leq 8/9$ is valid for
an arbitrary equation of state $p = p(\rho)$. Only a limited mass
can thus be contained within a given radius in general relativity.
The gravitational redshift $z = (1 - 2M/R)^{-\frac{1}{2}} - 1$ from
the surface of a static star cannot be higher than 2.

Many other explicit static perfect fluid solutions are known (we
refer to \cite{SK} for a list),
however, none of them can be considered as really ``physical''. Recently, the dynamical systems
approach to relativistic spherically symmetric static perfect
fluid models was developed which gives qualititative
characteristics of masses and radii (cf. work by C. Uggla et al.).

The most significant \textit{nonstatic} spacetime describing a
bounded region of matter and its external field is undoubtedly the
Oppenheimer-Snyder model of \textit{gravitational collapse of
a spherical star} of uniform density and zero pressure
(a ``ball of dust''). The model does not represent any new (local)
solution: the interior of the star is described by a part of a
dust-filled FLRW universe (cf. (\ref{FLRW})), the external region
by the Schwarzschild vacuum metric (cf. Eq. (\ref{Schwarzschild}),
Fig. 1).

Since Vaidya's discovery of a ``radiating Schwarzschild metric'',
null dust (``pure radiation field'') has been widely used as a
simple matter source. Its energy-momentum tensor, $T_{\alpha\beta}
= \varrho k_\alpha k_\beta$, where $k_\alpha k^\alpha = 0$, may be
interpreted as an incoherent superposition of waves with random
phases and polarizations moving in a single direction, or as
``lightlike particles'' (photons, neutrinos, gravitons) that move
along $k^\alpha$. The \textit{Vaidya metric} describing spherical
implosion of null dust implies that in case of a ``gentle'' inflow
of the dust, a naked singularity forms. This is relevant in the
context of the cosmic censorship conjecture (cf., e.g., \cite{Jo}).
\section{Cosmological models}
There exist important generalizations of the standard FLRW models
other than the above-mentioned Bianchi models, in particularly
those that maintain spherical symmetry but do not require 
homogeneity. The best known are the
\textit{Lema\^{\i}tre-Tolman-Bondi} models of inhomogeneous
\textit{universes} of pure dust, the density of which may
vary \cite{Kr}. (J. Gair recently generalized these models by
including anisotropic pressure and null dust.)

Other explicit cosmological models of principal interest involve, e.g., the G\"{o}del universe---a homogeneous, stationary spacetime with $\Lambda<0$ and incoherent rotating matter in which there exist closed timelike curves through every point; the Kantowski-Sachs solutions---possessing homogeneous spacelike hypersurfaces but (in contrast to the Bianchi models) admitting no simply-transitive G$_3$; and vacuum Gowdy models (``generalized Einstein-Rosen waves") admitting G$_2$ with compact 2-tori as its group orbits and representing cosmological models closed by gravitational waves. See [[entry Cosmology: Mathematical Aspects]] and Refs. \cite{SK}, \cite{BV}, \cite{JiBi}, \cite{HE}, \cite{Kr}.
\section*{See also:} 
General Relativity: Overview.
Stationary Black Holes. 
Asymptotic Structure and Conformal Infinity. 
Newtonian Limit of General Relativity.
Cosmology: Mathematical Aspects. 
Spinors and Spin-Coefficients. 
Spacetime Topology, Causal Structure and Singularities. 
Lie Groups and Lie Algebras. 
General Relativity: Experimental Tests.
\section*{Suggestions for further reading:} [1-14]

\end{document}